\documentclass[12pt]{article}
\usepackage{setspace}
\usepackage{subfigure}
\usepackage{graphicx}
\usepackage{caption}
\usepackage{amsmath}
\usepackage{amssymb}
\usepackage{cite}
\usepackage{color}
\usepackage{booktabs}
\usepackage{tabularx}
\usepackage{array}
\usepackage{makecell}
\usepackage{amsmath}
\usepackage{lineno}
\usepackage{upgreek}
\usepackage{svg}
\usepackage{geometry}
\usepackage{siunitx}
\usepackage{graphicx}

\usepackage[pagestyles]{titlesec}
\usepackage{xurl}  
\usepackage{sidecap}  
\usepackage{textcomp}  
\usepackage{xr-hyper} 
\usepackage{hyperref}
\usepackage{bm}
\usepackage{ulem} 

\if@Standard@Nature@refstyle%
  \usepackage[numbers,sort&compress]{natbib}%
  \def\bibfont{\reset@font\fontfamily{\rmdefault}\normalsize\selectfont}%
\fi%

\hypersetup{
	colorlinks=true,
	linkcolor=red,
	filecolor=blue,
	urlcolor=blue,
	citecolor=blue,
}

\usepackage[labelfont=bf,labelsep=period,justification=raggedright, font=small]{caption}

\makeatletter
\renewcommand{\@biblabel}[1]{\quad#1.}
\makeatother

\date{}
\begin{document}
\thispagestyle{empty}
\setcounter{page}{1}

\begin{center}
   {\Large
    \textbf{From Static to Dynamic Structures: Improving Binding Affinity Prediction with Graph-Based Deep Learning} \\
    }
    \vspace{0.5cm}
    Yaosen Min\,$^{1,\dagger}$,
    Ye Wei\,$^{1,\dagger,\ast}$,
    Peizhuo Wang\,$^{6,1,\dagger}$,
    Xiaoting Wang\,$^{2}$,
    Han Li\,$^{1}$,
    Nian Wu\,$^{1}$,
    Stefan Bauer\,$^{3}$,
    Shuxin Zheng\,$^{4}$,
    Yu Shi\,$^{4}$,
    Yingheng Wang\,$^{5}$,
    Ji Wu\,$^{5,\ast}$,
    Dan Zhao\,$^{1,\ast}$,
    and Jianyang Zeng$^{1,7,\ast}$
\end{center}

\begin{center}
\bf{1} Institute for Interdisciplinary Information Sciences, Tsinghua University, 100084, Beijing, China.\\
\bf{2} School of Medicine, Tsinghua University, 100084, Beijing, China. \\
\bf{3} Department of Intelligent Systems, KTH Stockholm, Sweden.\\
\bf{4} Microsoft Research Asia, 100080, Beijing, China. \\
\bf{5} Department of Electrical Engineering, Tsinghua University, 100084, Beijing, China. \\
\bf{6} School of Life Science and Technology, Xidian University, 710071, Xi'an, Shaanxi, China.\\
\bf{7} Present address: School of Engineering, Westlake University, 310030, Hangzhou, China. \\
$\dagger$ These authors contributed equally to this work. \\
$\ast$ All correspondence should be addressed to: weiye@mail.tsinghua.edu.cn, wuji$\_$ee@mail.tsinghua.edu.cn, zhaodan2018@tsinghua.edu.cn, and zengjy@westlake.edu.cn. \\
\end{center} 

\section*{Abstract}
\normalsize

Accurate prediction of protein-ligand binding affinities is an essential challenge in structure-based drug design. Despite recent advances in data-driven methods for affinity prediction, their accuracy is still limited, partially because they only take advantage of static crystal structures while the actual binding affinities are generally determined by the thermodynamic ensembles between proteins and ligands. One effective way to approximate such a thermodynamic ensemble is to use molecular dynamics (MD) simulation. Here, an MD dataset containing 3,218 different protein-ligand complexes is curated, and Dynaformer, a graph-based deep learning model is further developed to predict the binding affinities by learning the geometric characteristics of the protein-ligand interactions from the MD trajectories. \textit{In silico} experiments demonstrated that the model exhibits state-of-the-art scoring and ranking power on the CASF-2016 benchmark dataset, outperforming the methods hitherto reported. Moreover, in a virtual screening on heat shock protein 90 (HSP90) using Dynaformer, 20 candidates are identified and their binding affinities are further experimentally validated. Dynaformer displayed promising results in virtual drug screening, revealing 12 hit compounds (two are in the submicromolar range), including several novel scaffolds. Overall, these results demonstrated that the approach offer a promising avenue for accelerating the early drug discovery process.
\vspace{0.35cm}
\noindent \textbf{Key words:} Drug discovery, Binding affinity, Graph transformer, Molecular dynamics

\clearpage
\section{Introduction}\label{sec1}

Protein-ligand binding plays a key role in a wide range of biological processes, such as enzyme catalysis, signaling pathways, and the maintenance of cell structure~\cite{rube2022prediction, mcfedries2013methods}. Understanding the binding mechanisms of ligand molecules is essential for the design and development of novel drugs with high affinity and selectivity~\cite{swinney2011were, erlanson2016twenty}. Therefore, developing accurate computational methods for modeling protein-ligand binding and predicting the corresponding affinities is one of the most fundamental tasks in drug discovery.

Recent advances in computational methods have enabled protein-ligand binding affinity predictions through physics- and knowledge-based approaches~\cite{mobley2017predicting, kumar2021binding}. The molecular docking algorithms utilize coarse-grained physical models to predict and evaluate the binding conformation of small molecules. However, such methods usually incorporate expert knowledge and hand-tuned parameters, thus often leading to biased results~\cite{fischer2021decision}. All-atom molecular dynamics (MD) is another popular computational technique to investigate the structural and dynamical properties of biological systems. It is mainly based on Newtonian mechanics to reveal the full atomic detail at a fine spatial-temporal resolution, which is the basis of numerous binding affinity calculation approaches, such as MM/PB(GB)SA and alchemical free energy~\cite{frenkel2001understanding, valdes2021gmx_mmpbsa,song2020evolution}. Nevertheless, molecular dynamics simulation usually consumes a large amount of computational resources, which impedes its application to high-throughput virtual screening~\cite{liu2016applying}.

Data-driven approaches are another emerging trend for solving the binding affinity prediction task~\cite{lo2018machine}. Primarily benefiting from the rapidly growing number of experimental protein-ligand complexes, data-driven approaches have achieved a certain accuracy level through learning the binding modes directly from 3D structures. Nonetheless, current state-of-the-art methods often require sophisticated feature engineering while offering limited generalizability.~\cite{zhu2022assessment}. More importantly, various attempts, such as docking-based~\cite{verdonk2003improved,friesner2004glide,trott2010autodock,korb2009empirical}, feature-based~\cite{dittrich2018converging,mooij2005general,su2018comparative,wang2002further,ballester2010machine,wang2017improving,sanchez2021extended}, voxel-based~\cite{stepniewska2018development,jimenez2018k,zheng2019onionnet}, and graph-based~\cite{nguyen2021graphdta,yang2019analyzing,lim2019predicting,maziarka2020molecule,klicpera2020directional,song2020communicative,li2021structure} methods, mainly focus on tuning models on the X-ray crystallographic structures, which only capture the average conformational distributions of crystallized molecules and neglect their possible dynamic patterns. In fact, both proteins and ligands are not rigid structures, and they constantly interact with each other and the aqueous environment, following a specific conformation distribution. Such a conformation distribution in an equilibrium state is called a thermodynamic ensemble, and all the conformations collectively contribute to the affinity of the protein-ligand binding. Although such a thermodynamic ensemble can be obtained through MD simulation, large-scale MD trajectory datasets are still scarce due to the large amount of computational resources required. Recently, a number of noteworthy approaches have been proposed to address this issue. For instance, a recently developed molecular dynamics simulation framework, called TorchMD, leverages machine learning to enhance the empirical force fields in terms of efficiency and accuracy~\cite{doerr2021torchmd}. Another deep learning framework has been proposed to predict conformations in flexible regions of the target protein that rely on interactions with bound ligands~\cite{fernandez2020artificial}. Furthermore, other deep learning based approaches have also been explored to learn the MD trajectory features and then extend trajectories beyond the original MD-accessible timespans~\cite{fernandez2020deep,zheng2024predicting}. These existing works present promising approaches to overcome the current computational limitations of MD simulations. With this progress, recent efforts have been made as well to construct comprehensive MD simulation datasets through classical or machine learning force fields~\cite{siebenmorgen2023misato,priyakumar2023plas,Ariel2020,Doerr2021}. Nevertheless, few existing data-driven approaches have taken such thermodynamic ensembles into account to predict protein-ligand binding affinities~\cite{wu2022AdvancedScience}.

In light of this, we hereby explore the potential of integrating a data-driven deep learning model and MD simulated thermodynamic ensembles in improving the protein-ligand binding affinity prediction. We first curated a large-scale MD trajectory dataset, containing 3,218 different protein-ligand complexes, based on the PDBBind dataset~\cite{wang2004pdbbind}. Then, we introduced a graph transformer framework, named Dynaformer, which was trained based on this MD trajectory dataset. To capture the binding modes between proteins and ligands, Dynaformer utilizes a roto-translation invariant feature encoding scheme, taking various interaction characteristics into account, including interatomic distances, angles between bonds, and various types of covalent or non-covalent interactions. Compared to other existing methods, Dynaformer showed superior scoring and ranking power on the CASF-2016 benchmark dataset~\cite{su2018comparative}. Through several case studies, we also discussed the underlying thermodynamic binding mechanisms that may contribute to the performance improvement achieved by Dynaformer. Furthermore, we experimentally validated several novel hits identified by Dynaformer against heat shock protein 90 (HSP90), an attractive drug target against cancer. We demonstrated that our model can be applied to a real-world hit discovery task and serve as an effective tool to accelerate the drug discovery process.

\section{Results}\label{sec:results}

\subsection{The Thermodynamic Ensemble Determines the Binding Affinity}\label{subsec2.1}

\begin{figure}[htp] 
\centering
\includegraphics[width=1.0\textwidth]{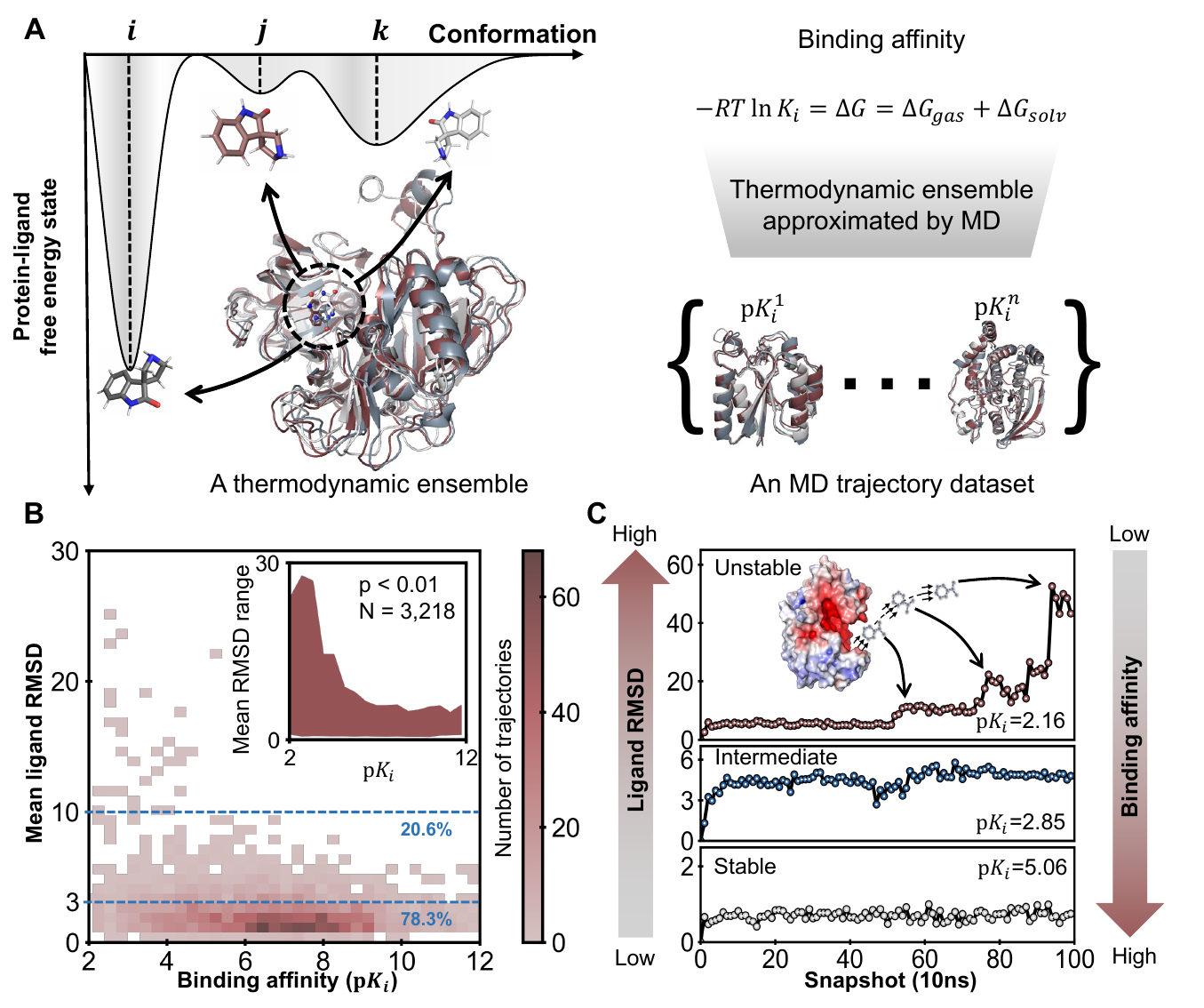}
\caption{\textbf{An illustrative overview of the thermodynamic ensemble and the MD trajectory dataset.} \textbf{A}) The conformations in a thermodynamic ensemble generally follow a specific distribution. In principle, the binding affinity is determined by the properties of the thermodynamic ensemble rather than a single conformation. Based on this fact, we built a dataset containing various thermodynamic ensembles sampled by MD simulations. More details can be found in the main text. \textbf{B}) Relationship between the mean ligand RMSD and the binding affinity in the MD trajectory dataset. The mean ligand RMSD is calculated as the average distance deviation of the ligand from its original position in the crystal structure during the simulation. The heatmap and the mean RMSD range in the inset indicate that the ligand dynamics are related to binding affinities, i.e., the higher the mean ligand RMSD, the lower the binding affinity. \textbf{C}) Representative examples of ligand RMSD and their binding affinities, including unstable (PDB ID: 4y3j), intermediate (PDB ID: 3udh), and stable (PDB ID: 2yge) MD trajectories. In the unstable case, the ligand RMSD increased sharply at the 60th snapshot, indicating that the ligand left the binding site. The ligand with intermediate stability remained relatively flexible but stayed near the binding site. In the stable case, the ligand stuck tightly to the binding pocket, yielding low ligand RMSDs.}\label{fig:gibbs}
\end{figure}

A thermodynamic ensemble of a protein-ligand system represents a distribution of complex conformations in equilibrium. Theoretically, these conformations, with varying probabilities of occurrence, collectively depict the free energy state of the system~\cite{laio2008metadynamics, christ2010basic, wei2016protein}. Therefore, conformation snapshots sampled from the thermodynamic ensemble can approximate the distribution of conformation space and further reflect the free energy state of ligand binding. In other words, an MD trajectory can provide more information about ligand binding than a single static structure (Figure~\ref{fig:gibbs}\textbf{A}). More specifically, the binding affinity $K_i$ is determined by the Gibbs free energy change $\Delta G$ of the binding process~\cite{du2016insights}, which can be defined as:
\begin{equation}
\begin{split}
    -RT\textrm{ln} K_i = \Delta G = \Delta G_{gas} + \Delta G_{solv},
\end{split}
\label{eq0}
\end{equation}
where $R$ represents the molar gas constant, $T$ represents the temperature, $\Delta G_{gas}$ stands for the binding free energy between the pair of protein and ligand in the gas phase, and $\Delta G_{solv}$ stands for the solvation free energy difference between the protein-ligand complex and the sum of stand-alone protein and ligand systems. Both $\Delta G_{gas}$ and $\Delta G_{solv}$ can be represented by the ensemble-averaged energy terms related to the atomic features derived from each snapshot (see Experimental Section for more details). Therefore, in principle, with sufficiently sampled snapshots from an MD simulation, the binding affinity can be expressed as a function of the atom features of the snapshots.

Based on the above principle, training deep learning models to model the energy terms using the thermodynamic ensembles sampled from MD simulations can enable one to predict the protein-ligand binding affinities. Here, we first built a large-scale dataset containing MD simulations of 3,218 protein-ligand complexes based on experimental structures derived from the Protein Data Bank (PDB)~\cite{berman2000protein}. All of the complexes were further filtered based on the PDBBind dataset~\cite{wang2004pdbbind}, which consists of protein and small molecule complex structures with known binding affinities. In this study, we performed a 10-nanosecond (ns) simulation for each complex and sampled 100 snapshots from each simulation to characterize the conformational space. All snapshots were collected to derive the MD trajectory dataset.

To gain comprehensive insight into the relationships between complex stabilities and corresponding binding affinities, we also performed a deep analysis of our MD trajectory dataset. More specifically, we first derived the relative position movement of each ligand during the simulation by calculating the average root mean squared deviation (RMSD) between its initial and current positions over 100 snapshots at the binding site. Then, we examined the association between the mean RMSDs of ligands and experimentally measured binding affinities. Our analysis revealed that conformations from MD simulations exhibited varying levels of stability, and the stabilities of ligands were roughly inversely correlated with binding affinities (Figure~\ref{fig:gibbs}\textbf{B}). More specifically, the upper bound for the mean RMSD of the ligand decreased as the binding affinity increased. In total, 78.3\% of trajectories in the dataset exhibited high stability, with mean RMSDs of ligands smaller than 3\r{A}; 20.6\% of trajectories displayed intermediate flexibility, where certain atoms or functional groups in the ligands were flexible but did not leave the original binding sites, with mean RMSDs of ligands between 3\r{A} and 10\r{A}; and 1.1\% of complexes were unstable, with mean RMSDs of ligands greater than 10\r{A}, exhibiting a high level of ligand flexibility. In the unstable cases, the binding affinities tended to be low, and ligands even left the original binding pockets and entered the solvent. Several typical examples are shown in Figure~\ref{fig:gibbs}\textbf{C}. In the unstable example (PDB ID: 4y3j) with the lowest binding affinity, the ligand left the binding site shortly after the simulation began, while in the intermediate (PDB ID: 3udh) and stable (PDB ID: 2yge) examples, the ligands remained in the binding sites, displaying relatively higher binding affinities. This preliminary data analysis validated that ligand dynamics can provide extra information with respect to the corresponding binding affinity. The Dynaformer predictions on the snapshots of the trajectories, in turn, provide additional evidence for this observation. In addition, the variances of Dynaformer predictions were higher for weaker binders, thus indicating lower confidence levels. Therefore, our MD trajectory dataset contains more information than its static crystal structure counterparts and thus should be able to contribute to binding affinity prediction (Figure~S3, Supporting Information).

\subsection{Virtual Screening Pipeline with Dynaformer}\label{sec:framework}

\begin{figure}[htp]
\centering
\includegraphics[width=0.9\textwidth]{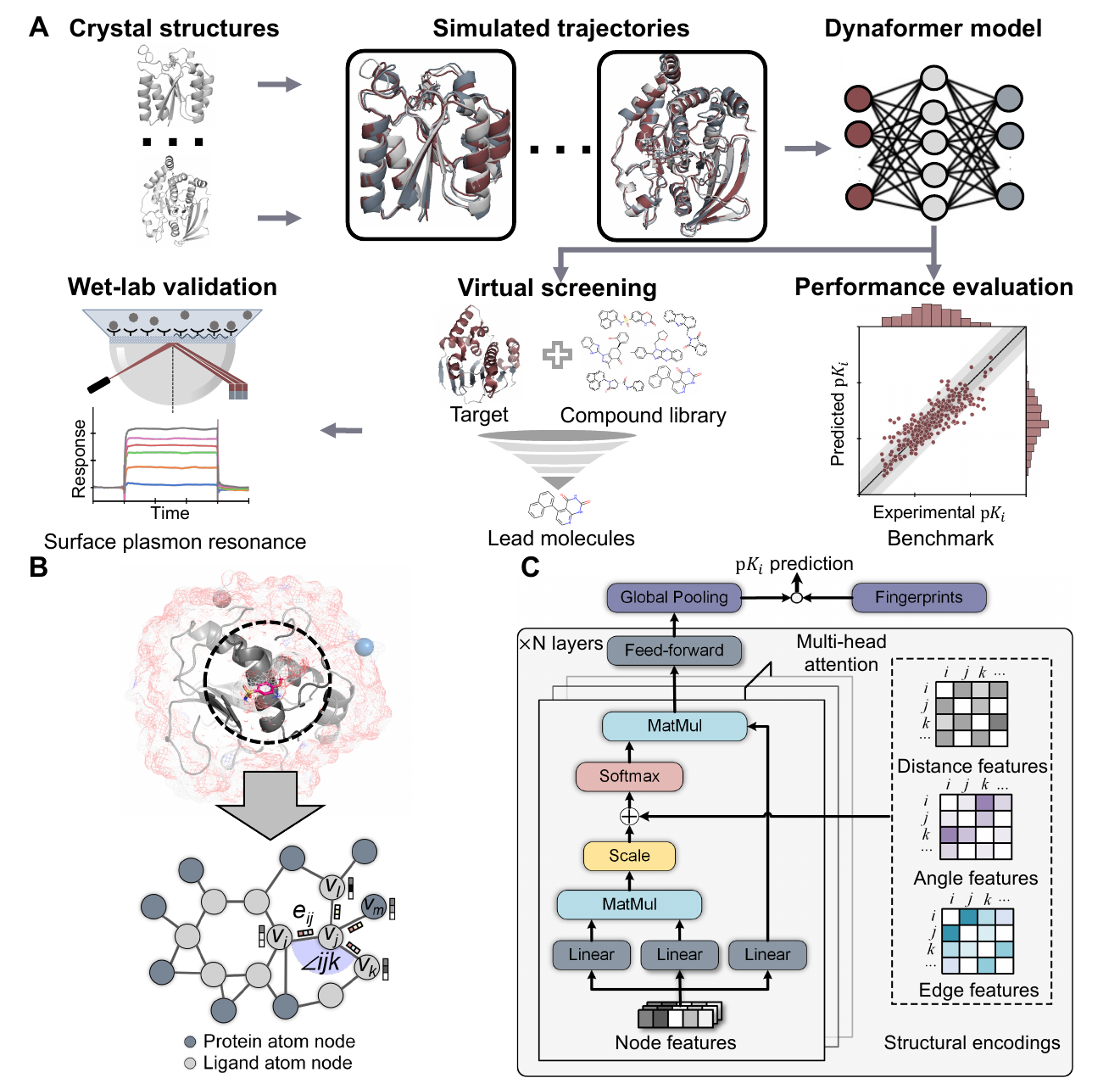}
\caption{\textbf{Illustration of our virtual screening pipeline with Dynaformer.} \textbf{A}) Hit discovery through our deep learning model Dynaformer based on an MD trajectory dataset. First, crystal structures of protein-ligand complexes are collected to form an MD trajectory dataset. Then, the simulated trajectories are utilized for training the binding affinity prediction model. Next, the trained model is evaluated against a benchmark dataset. The model with the best performance is then used as a scoring function for the virtual screen of the compound candidates. Finally, the wet-lab experiment validates the hit molecules for further hit-to-lead optimization. \textbf{B}) An example snapshot of protein-ligand from the MD trajectory dataset and its graph representation. In order to create such a graph representation to feed into the model, atoms from both the ligand and the protein are chosen based on their proximity to the ligand within a specified distance cutoff. The nodes and the edges represent the atoms and their covalent or non-covalent interactions, respectively. \textbf{C}) The architecture of Dynaformer. The node features (i.e., atomic features) are first fed into a multi-head attention module. The structural encodings, including distance, angle, and edge features, are encoded and added as the attention bias. After the final layer, the feature representations of the graph are globally pooled and fused with the pre-calculated fingerprints, which are knowledge-based features capturing structural and chemical properties. The fused representations are then used for the final prediction of the binding affinity. Abbreviations in the figure: Linear, linear layer; Feed-forward, feed-forward neural network; MatMul, matrix multiplication.}\label{fig:intro}
\end{figure}

An overview of our virtual screening pipeline using Dynaformer is shown in Figure~\ref{fig:intro}\textbf{A}, including the MD trajectory dataset preparation, model training, performance evaluation, virtual screening, and wet-lab validation. To feed data into our deep learning model, snapshots from MD trajectories are converted into graphs representing the ligand binding structures. Each snapshot is converted into a graph containing nodes and edges that represent features of atoms and their relationships, such as chemical bonds and non-covalent interactions (Figure~\ref{fig:intro}\textbf{B}). A graph includes atoms and the interatomic interactions from both the ligand and the protein within a certain distance cutoff. After preparing the MD trajectory dataset, we propose a novel graph transformer architecture, named Dynaformer (Figure~\ref{fig:intro}\textbf{C}), to fully take advantage of the characteristics of molecular dynamics from the dataset and improve the predictions of protein-ligand binding affinities.

Inspired by Graphormer, Dynafomer extends it by integrating interatomic interaction features that encapsulate the spatio-temporal relationships inherent in MD trajectories~\cite{ying2021transformers}. To capture the intricate non-covalent interaction features, Dynaformer introduces additional attention bias terms to encode the structural information regarding the interatomic distances and angles between these bonds. In addition, the Gaussian basis function (GBF) is used for structural encodings, which is a common form of energy function in computational chemistry to map structural information to energy contributions (Equation~\ref{eq:gbf} in Experimental Section)~\cite{bartok2010gaussian}. In this way, from the graph representation of a protein-ligand complex, we are able to model the protein-ligand interactions of a snapshot from an MD trajectory. Moreover, three fingerprints from ECIF~\cite{sanchez2021extended}, RF-score~\cite{ballester2010machine}, and GB-score~\cite{rayka_firouzi_2022} are used to model longer-range interactions that may affect ligand binding beyond the distance cutoff of the graph representation. These fingerprints are fused into the final linear layer.

As mentioned previously, MD simulations are notoriously time-consuming, making MD trajectory data rarely used in high-throughput virtual screening. To ensure practical applicability and accurate prediction in realistic scenarios, we employed a pretraining-finetuning strategy in our framework. More specifically, we first pretrained Dynaformer on the graphs derived from the MD trajectory dataset through predicting the corresponding binding affinities. All graphs from the same MD trajectory shared the same binding affinity label in pretraining, suggesting an equal contribution to the binding affinity from each snapshot within a thermodynamic ensemble. The time order of snapshots in MD trajectories was therefore discarded. This setting allowed the model to infer binding affinity from a single snapshot, encapsulating information such as ligand interactions and atomic flexibility from the whole trajectory. Then, for the benchmarking test and the virtual screening applications, the pretrained model was further finetuned on the PDBBind dataset. This finetuning stage was designed to align the upstream MD pretraining process with downstream virtual screening applications to achieve better performance. Furthermore, it can eliminate the need of time-consuming MD simulations in the inference stage. More details about the Dynaformer architecture and the pretraining-finetuning strategy can be found in the Experimental Section. After finetuning, the Dynaformer model can serve as a scoring function for novel hit discovery in a real-world virtual screening application. In detail, the docking structures between the target of interest and the molecules from a compound library are fed into the model to predict the binding affinity. The top-ranked molecules are then selected as the hit molecules and further investigated in the downstream wet-lab experiments (Figure~\ref{fig:intro}\textbf{A}).

\subsection{Molecular Dynamics Data Improves Binding Affinity Prediction}\label{subsec2.3}

\begin{figure}[htp]
\centering
\includegraphics[width=0.85\textwidth]{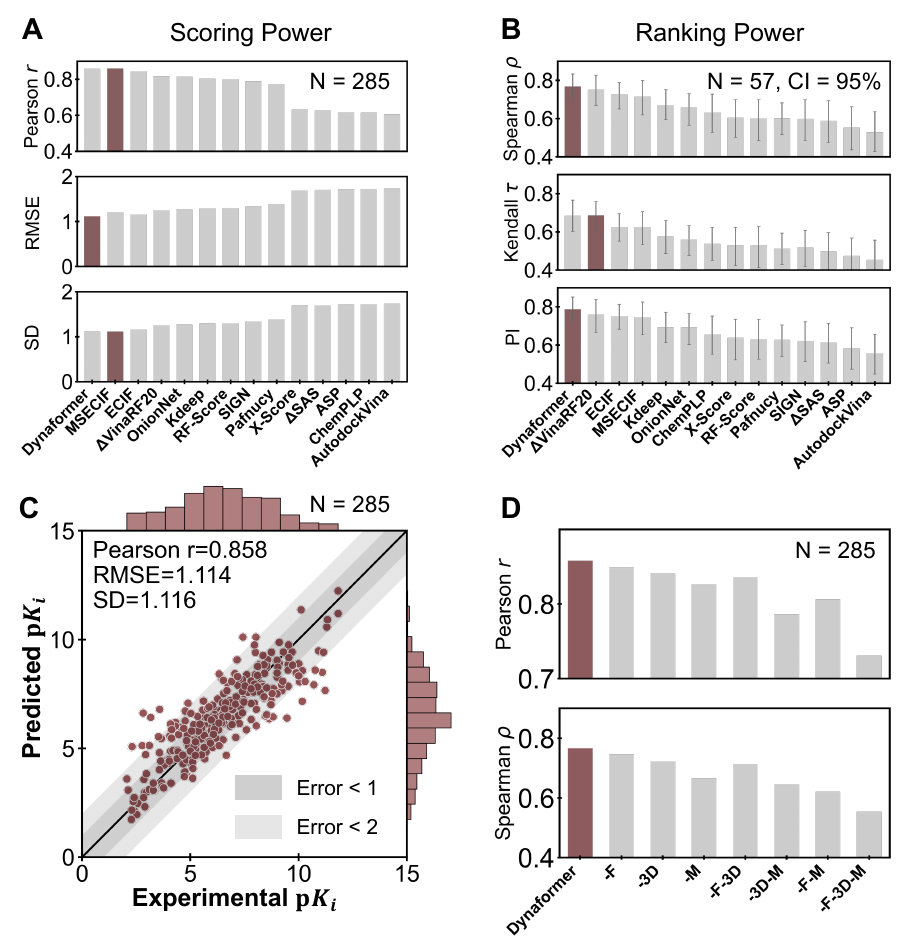}
\caption{\textbf{Performance evaluation of binding affinity prediction.} \textbf{A}) Scoring power evaluation on the 285 CASF-2016 complexes. The scoring power illustrates the inter-target performance of how well the predictions correlate with the experimentally measured binding affinities. Pearson coefficient (Pearson $\bm{r}$), root mean squared error (RMSE), and standard deviation (SD) are used as evaluation metrics. \textbf{B}) Ranking power evaluation on the 57 classes of target proteins from CASF-2016, in which each target bound to five different ligands. The goal of evaluating the ranking power is to assess the intra-target performances of scoring functions in terms of how accurately they predict the relative order of molecules. Evaluation metrics are measured in terms of the average values of Spearman's coefficient (Spearman $\rho$), Kendall's coefficient (Kendall $\tau$), and predictive index (PI) on each target. \textbf{C}) A detailed scatter plot of the Dynaformer predicted $\mbox{p}K_i$ and experimentally measured $\mbox{p}K_i$ values of the 285 protein-ligand complexes from CASF-2016, demonstrating the effectiveness of Dynaformer in predicting binding affinity with high correlation and low prediction bias. \textbf{D}) Ablation studies of Dynaformer. Here, -F indicates a modified version of Dynaformer without any pre-calculated fingerprints; -3D stands for a modified version of Dynaformer without the structural encoding module; -M stands for a modified version of Dynaformer that is trained without MD trajectories, i.e., only crystal structures are used for training; The results showed consistent improvements in both scoring and ranking power by incorporating MD data.}\label{fig:casf-eval}
\end{figure}

To evaluate the performance of Dynaformer on binding affinity prediction, we tested it on the CASF-2016 benchmark dataset, which consisted of 285 protein-ligand complexes covering 57 different target classes~\cite{su2018comparative}. Following the CASF-2016 evaluation protocols, we compared the scoring and ranking power of Dynaformer to a variety of baseline methods, including docking-based~\cite{verdonk2003improved,friesner2004glide,trott2010autodock,korb2009empirical}, feature-based~\cite{dittrich2018converging,mooij2005general,su2018comparative,wang2002further,ballester2010machine,wang2017improving,sanchez2021extended, shiota2023multi}, voxel-based~\cite{stepniewska2018development,jimenez2018k,zheng2019onionnet}, and graph-based~\cite{nguyen2021graphdta,yang2019analyzing,lim2019predicting,maziarka2020molecule,klicpera2020directional,song2020communicative,li2021structure} methods. In this evaluation, the same PDBBind dataset for finetuning Dynaformer was utilized for the training of all baseline methods, with the exception of ECIF~\cite{sanchez2021extended} and MSECIF~\cite{shiota2023multi}, as they were trained on a PDBBind subset as reported in the original studies. Moreover, the prediction results of docking-based and other feature-based methods were obtained from the CASF-2016 benchmark dataset. This experimental setting guaranteed a fair comparison of different methods. The scoring power of a model refers to its ability to accurately predict binding affinities, which can be measured using the metrics Pearson's correlation coefficient (Pearson $\bm{r}$), root-mean-square error (RMSE), and standard deviation (SD), while the ranking power of a model indicates its ability to correctly rank the relative order of binding ligands, which can be measured using the metrics Spearman's coefficient (Spearman $\bm{\rho}$), Kendall's coefficient (Kendall $\bm{\tau}$), and predictive index (PI). The scoring and ranking performances of different methods are shown in Figure~\ref{fig:casf-eval}\textbf{A} and \textbf{B}, respectively. The evaluation results showed that Dynaformer outperformed all the baseline methods. In addition, the prediction scatter plot in Figure~\ref{fig:casf-eval}\textbf{C} revealed a high correlation with a Pearson $\bm{r}$ of 0.858 and a low prediction bias with an RMSE of 1.114 between the Dynaformer predicted and experimentally measured binding affinities. 

We next carried out comprehensive ablation studies on Dynaformer, to validate the efficacy of its specific designs (Figure~\ref{fig:casf-eval}\textbf{D}). We first introduced multiple variations of the Dynaformer model, such as -F (i.e., Dynaformer without the fingerprints), -3D (i.e., without the structural encoding module), and -M (i.e., without pretraining on the MD dataset). Then, we compared the prediction performances of Dynaformer and its variations using Pearson $\bm{r}$ and Spearman $\bm{\rho}$. Dynaformer achieved superior performances in comparison with its variations, thus indicating the effectiveness of the individual modules employed in Dynaformer. More specifically, as expected, the pretraining on the MD trajectory data played the most significant role in enhancing the scoring and ranking power of Dynaformer. In addition, the 3D structural encoding module, which encodes distance and angle information using the GBF module, also played a crucial role in improving the prediction performance. Moreover, incorporating the pre-calculated fingerprints also yielded a performance gain, probably due to the increased receptive field. It is noteworthy that the architecture of Dynaformer significantly contributes to the performance of binding affinity prediction. Even without the pretraining on the MD dataset (i.e., the -M variation in Figure~\ref{fig:casf-eval}\textbf{D}), the performance of Dynaformer remained comparable with the best of baseline methods that were trained with the same crystal structures. Feature-based methods, such as ECIF, while capable of capturing interaction features of protein-ligand complexes, were unable to model the geometric topologies of interactions. Voxel-based and graph-based methods that utilize 3D convolutional neural networks or graph neural networks, while capable of modeling geometric topologies of interactions, had limited capacity in capturing remote interactions due to the locality nature of the employed networks. This highlighted the strength of more generalized graph transformers, such as Dynaformer, which can capture both geometric relationships and remote interactions, thus boosting the prediction performance. 


To demonstrate that incorporating the pretraining on the MD dataset is necessary, we further investigated whether Dynaformer can capture the intrinsic binding-related patterns from the MD trajectory dataset, such as the entropy feature, the enthalpy feature, and the subtle differences in protein-ligand interactions. More specifically, we performed three case studies from the test set of CASF-2016, where Dynaformer offered more accurate predictions in comparison with baseline methods. We employed ECIF and SIGN as baselines for comparison in these case studies, which were fingerprint-based and graph-based binding affinity prediction models trained on static crystal structures, respectively. 

\begin{figure}[htp]
\centering
\includegraphics[width=0.86\textwidth]{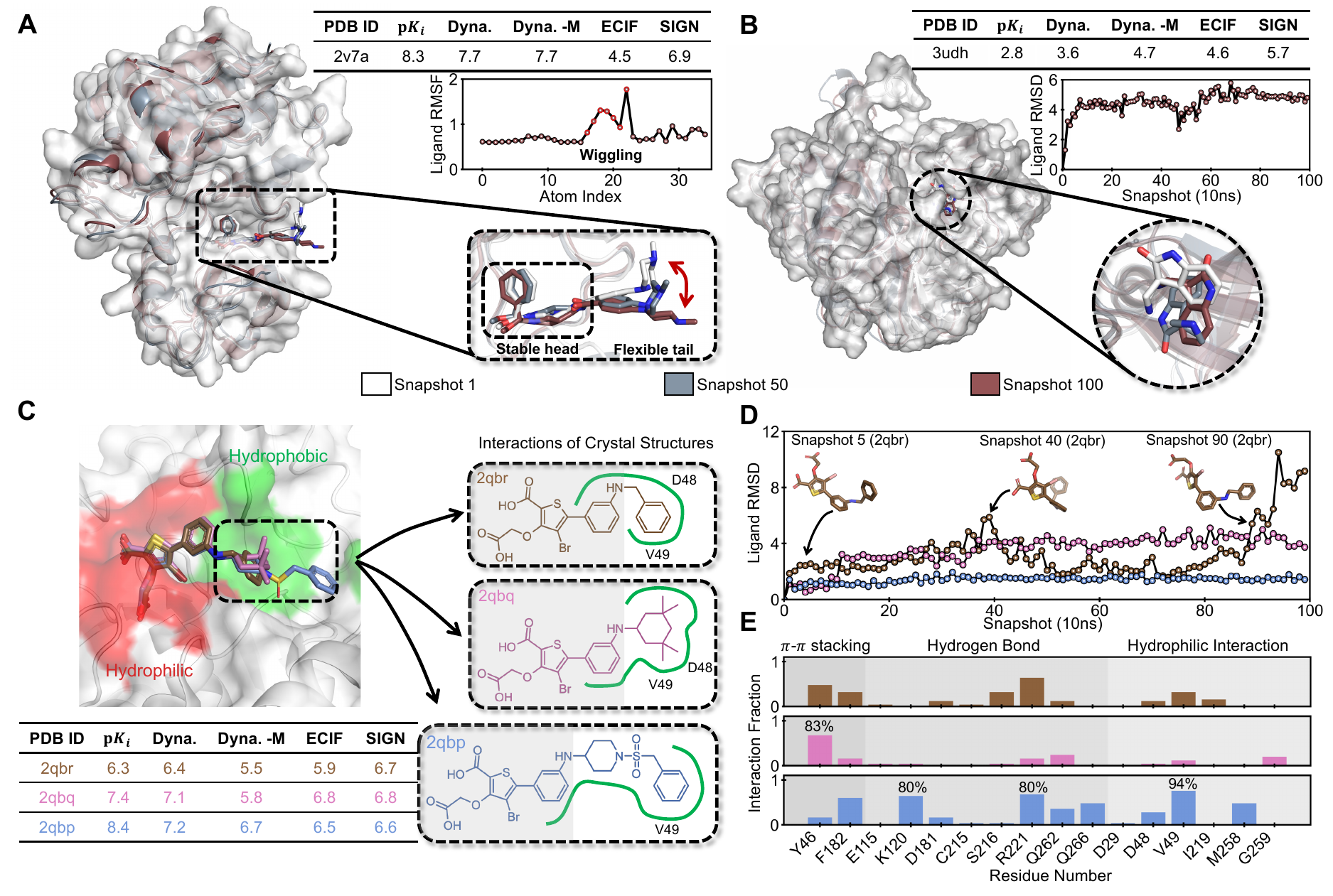}
\caption{\textbf{Illustrative examples of Dynaformer trained with MD trajectories improves binding affinity prediction.} \textbf{A}) An example (PDB ID:2v7a) of improved affinity prediction through incorporating additional entropic information. The head of the ligand is tightly bound in the pocket, resulting in a significant favorable enthalpy change. The random movement of the solvent-exposed ligand tail contributes to a favorable entropy change during ligand binding. \textbf{B}) An example (PDB ID: 3udh) of improved affinity prediction through better protein-ligand interaction modeling. With a rigid structure and no rotatable bonds, the binding affinity relies primarily on the enthalpy changes from binding site interactions. Despite a few favorable interactions, subpockets around the ligand remain unoccupied. Therefore, the ligand moves away from its initial position during the simulation, leading to low binding affinity. The ability of Dynaformer to model these interactions results in better prediction. \textbf{C}) An example of ranking ligands with subtle interaction differences. The three ligands (PDB IDs: 2qbr, 2qbq and 2qbp) have the same scaffold (the shaded area) but different tail structures. Dynaformer better addresses activity cliffs by differentiating between ligands sharing the same scaffold but with varying binding affinities. \textbf{D}) Conformational stability of the three ligands shown in \textbf{C}. The 2qbp ligand exhibited high stability during the simulation, while the 2qbq ligand partially left the original binding site after 1 ns and the 2qbr ligand left the binding site after 9 ns. The varying binding affinities of these ligands can be attributed to the hydrophobic interactions with the second subpocket on the right side shown in \textbf{C}. \textbf{E}) The interaction fraction analysis of the three ligands shown in \textbf{C}. The percentages of simulation snapshots with different protein-ligand interactions are shown. Higher binding affinity corresponded to more stable interactions (e.g., $\pi-\pi$ stacking, hydrogen bond and hydrophilic interaction) observed during the simulation. With such interaction features and better modeling capability, Dynaformer can provide more accurate ligand rankings. ECIF and SIGN are selected as baselines for comparison in \textbf{A}-\textbf{D}.}\label{fig:dynamics}
\end{figure}

In the first case, we showed that Dynaformer can predict the binding affinity more accurately through learning the ligand flexibility, i.e., the additional entropic information derived from the MD trajectory. In the protein-ligand complex structure shown in Figure~\ref{fig:dynamics}\textbf{A} (PDB ID: 2v7a), which is a ligand bound to the T315I Abl kinase domain~\cite{modugno2007crystal}. As can be seen in the zoomed-in figure, the head of the ligand (i.e., the pyrrolo[3,4-c]pyrazole group) is tightly buried in the binding pocket, whose contacts with the protein are highly conserved during the MD simulation. In detail, the carbonyl oxygen of residue E316 and the amide nitrogen of residue M318 are found to interact with the two nitrogen atoms of the pyrrolopyrazole scaffold, and the nitrogen of the amide group forms a hydrogen bond with the carbonyl oxygen of residue M318. In addition, the benzyl group is also involved in hydrophobic interactions with residue L370. These interactions on the head of the ligand lead to a considerably favorable enthalpy change due to the formation of non-covalent interactions. In addition, the tail of the ligand (i.e., the N-methylpiperazine group) is exposed to the solvent and wiggles randomly, which is also entropically favorable for ligand binding. This wiggling can be confirmed from the MD trajectories, as shown in the root mean square fluctuation (RMSF) plot in Figure~\ref{fig:dynamics}\textbf{A}, which measures the fluctuation of an atom around its average position. Consequently, this ligand showed a strong binding affinity, i.e., $\mbox{p}K_i = 8.3$. In the prediction results, ECIF and SIGN underestimated the binding affinity by a relatively large margin, while Dynaformer predicted the binding affinity more accurately. In this case, both the crystal structure and MD trajectory data contain enthalpy information. But the MD trajectory data includes additional entropic information, i.e., the features of ligand flexibility, which play a crucial role in the binding affinity. Through incorporating this extra information, Dynaformer was able to achieve better prediction accuracy.

In the second case, we show that Dynaformer can more accurately predict the binding affinity by modeling the enthalpy change, which is the energy term caused by the ligand-protein interactions. As demonstrated in Figure~\ref{fig:dynamics}\textbf{B}, the aspartyl protease $\beta$-secretase (BACE) is in complex with a compound in the spiropyrrolidine scaffold (PDB ID: 3udh), which was screened from a fragment library~\cite{efremov2012discovery}. Unlike the molecule shown in the first case, the ligand here has a rigid structure, i.e., a single conformational state and zero rotatable bonds. Therefore, its binding affinity depends mainly on the enthalpy change caused by the interactions with the binding site. In the determined structure, the pyrrolidine nitrogen is located between the two catalytic acids on the side chain of residues D32 and D228, forming stable hydrogen bonds. In addition, the oxindole and the phenyl ring next to the pyrrolidine ring exhibit strong complementarity in terms of the occupancy of the binding site. However, the adjacent subpockets are still left unoccupied, as it was the primary optimization direction as discussed in the original work~\cite{efremov2012discovery}. As a result, the binding affinity of the ligand was as low as $\mbox{p}K_i = 2.8$. In our MD trajectories, the ligand gradually moved away from its initial position after a few nanoseconds, as evidenced by an average ligand RMSD of over $5\mbox{\r{A}}$. Such behavior is typically associated with a weak binding affinity, despite some favorable protein-ligand interactions. Our test validated that the Dynaformer can learn such knowledge through training based on the MD trajectories, thus providing a more accurate prediction ($\mbox{p}K_i = 3.6$). Similar behaviors were also observed in other protein-ligand complexes with low binding affinities, such as the complexes with PDB IDs 4y3j and 3gv9.

With the third case shown in Figure~\ref{fig:dynamics}\textbf{C}, Dynaformer demonstrated its ability to distinguish ligands with subtle differences in protein-ligand interactions, thus allowing it to tackle the “activity cliff”, where similar ligands have distinct binding affinities. Here, three ligands with a thiophene scaffold bind to PTP1B (protein tyrosine phosphatase 1B) with varying potency, in which the ligand of PDB ID 2qbp showed ten times stronger binding affinity compared to that of PDB ID 2qbq, and the ligand of PDB ID 2qbq showed ten times stronger binding affinity compared to that of PDB ID 2qbr~\cite{wilson2007structure}. 2-phenylthiophene, as the structural core of these three compounds, adopts an identical binding pose to occupy the active site, and the tail groups enter another hydrophobic subpocket. The ligand RMSD plots in Figure~\ref{fig:dynamics}\textbf{D} indicated that the ligand from 2qbp exhibited high conformational stability at the binding site (mean ligand $\mbox{RMSD}<1$\r{A}), which was likely due to the highly conserved $\pi$-$\pi$ stacking with residue F182 and hydrogen bonds with residues K120 and R221. On the other hand, even though the other two ligands shared similar interactions as 2qbp, the ligand of PDB ID 2qbq partially left the original binding site after 1 ns of simulation, and the structural core of the ligand of PDB ID 2qbr finally left the binding site after around 9 ns of simulation. Figure~\ref{fig:dynamics}\textbf{E} displays the percentage of snapshots that key interactions were observed during the simulation. Such statistical information was another piece of evidence showing how conserved the interactions were. The occupancy of the hydrophobic subpocket by the different tail groups decreased progressively with each modification in the ligand structure, from 2qbp to 2qbq and finally to 2qbr. Therefore, these structurally similar ligands provided distinct binding strengths, with $\mbox{p}K_i$ = 8.4, 7.4, and 6.3 for 2qbp, 2qbq, and 2qbr, respectively. Through the rich interaction information learned from the MD trajectories and its ability to model such interactions, Dynaformer achieved a better ranking result compared to those of baselines ECIF and SIGN (Figure~\ref{fig:dynamics}\textbf{C}) and thus may provide more reliable guidance toward the downstream hit discovery task. 

In summary, the above detailed analyses showed that the MD trajectories may provide useful insights in understanding the protein-ligand binding affinities, and our Dynaformer model was capable of capturing such rich features for achieving better prediction results. 

\subsection{Hit discovery for the HSP90 target}\label{subsec2.4}

To verify the applicability of Dynaformer in realistic scenarios such as hit discovery in the early drug discovery process, we applied Dynaformer to discover potent hit compounds through scoring the docked protein-ligand structures. The original docking score often contains many false positives among the top-scoring candidates~\cite{lyu2019ultra}. Consequently, numerous experiments are often required to identify the true hit compounds. In this study, we demonstrate that Dynaformer can more efficiently deliver hit compounds by scoring docked poses. In particular, we selected heat shock protein 90 (HSP90) as our target protein, which is a vital chaperone protein involved in the important biological pathways of many refractory diseases, including cancer, neurodegenerative diseases, and viral infections. To inhibit the function of HSP90, a common strategy is to design small molecules targeting its ATP binding pocket to suppress the ATPase activity in the N-terminal domain. Here, our training data from the PDBBind dataset only included 18 protein-ligand structures associated with ATPase activity, with $\mbox{p}K_i$ values ranging from 3.84 to 8.32. All of these structures can be categorized as known inhibitor scaffolds summarized in the literature~\cite{janin2010atpase, li2019heat}.

Next, we performed docking on HSP90 against the ChemBridge DIVERSet-EXP library, which contains 50,000 small molecules that cover a diverse pharmacophore space. The docked conformers were obtained using Autodock Vina, a widely used open-source docking software~\cite{eberhardt2021autodock, trott2010autodock}. The reference protein-ligand structure (PDB ID: 2xdl) was chosen from the test set, CASF-2016, to identify the binding pocket. Then, the docked poses were used to score and rank the molecules based on the Dynaformer predictions. After that, the top-ranked compounds were filtered and visually inspected for prioritization, following the approach and criteria described in the literature~\cite{bender2021practical, fischer2021decision} (see Experimental Section for details). To ensure the fairness of prioritization, we created a pool consisting of molecules among the top 10\% from both Dynaformer and the original scoring function in Autodock Vina. From the pool, 20 molecules were chosen without knowing their rankings. Finally, surface plasmon resonance (SPR) experiments were conducted for each of the 20 molecules to measure the inhibition constant $K_i$. In total, 12 of the 20 molecules had measurable binding affinities.

\begin{figure}[htp]
\centering
\includegraphics[width=0.9\textwidth]{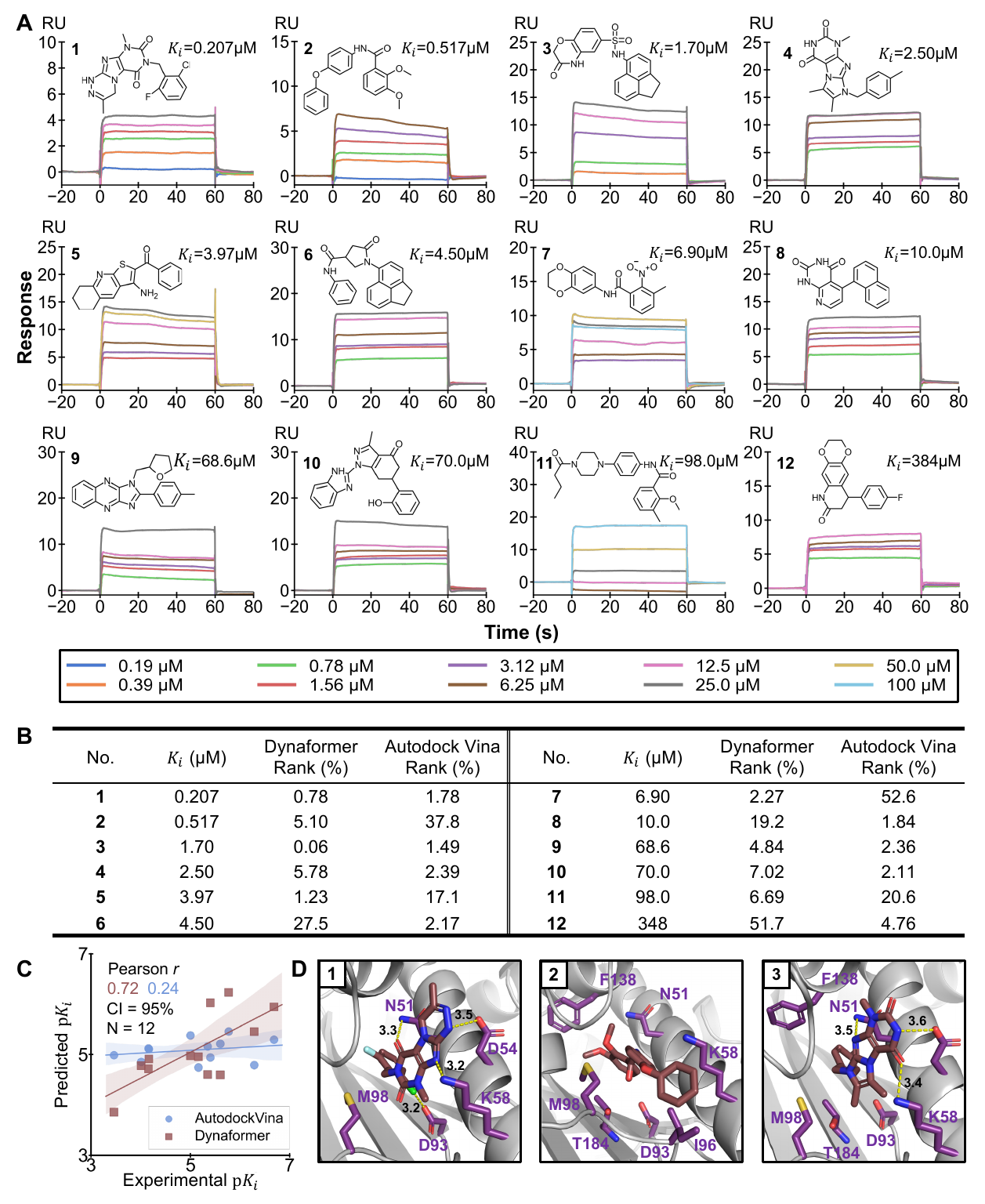}
\caption{\textbf{Virtual screening against the target HSP90 and the corresponding experimental validation.} \textbf{A}) Structures, experimentally measured $K_i$ values, and the SPR sensorgrams of the top 12 molecules with measurable binding affinities from the virtual screening. \textbf{B}) Detailed $K_i$ values of the selected 12 molecules and their ranking statistics by Dynaformer and Autodock Vina. To the best of our knowledge, compounds 1, 3, 4, 6, and 12 show promise as novel candidates for hit-to-lead optimization. \textbf{C}) The correlations of the predictions by Dynaformer and Autodock Vina predictions versus the experimental $K_i$ values of the 12 compounds. \textbf{D}) The possible binding modes of the top three compounds, in which the interacting residues are consistent with the previous studies. Overall, the wet-lab validations demonstrated that Dynaformer can effectively identify hit compounds with favorable binding affinities and novel scaffolds.}\label{fig:exp}
\end{figure}

Figure~\ref{fig:exp}\textbf{A} displays the chemical structures, experimentally measured $K_i$, and SPR sensorgrams of the 12 molecules that showed measurable binding affinities. The predicted and experimentally measured $K_i$ values are summarized in Figure~\ref{fig:exp}\textbf{B}, for comparing the prediction power of Dynaformer and Autodock Vina. The only difference here between Dynaformer and Autodock Vina was the scoring function used for binding affinity prediction, and the input docking structures remained the same. The results demonstrate that Dynaformer is able to predict highly potent molecules more effectively than the scoring function used in Autodock Vina, and the Pearson $\bm{r}$ coefficient indicates that the performance of Dynaformer (0.72) outperforms Autodock Vina (0.24) by a large margin (Figure~\ref{fig:exp}\textbf{C}). More specifically, Dynaformer ranked the top three compounds in the library at 0.78\%, 5.10\%, and 0.06\%, which is much higher than Autodock Vina's ranking. In addition, the possible binding modes of the top three compounds were consistent with the known binding patterns from previous studies of rational drug design (Figure~\ref{fig:exp}\textbf{D}). For example, previous investigations revealed that one) the hydrogen bond networks between inhibitors and residues D93, T184, and K58 are crucial for stable binding, and two) enhanced affinity and selectivity can be achieved when the lipophilic bottom of the pocket (including residues M98, F138, Y139, V150, and V186) is occupied by the aromatic moiety of the ligand~\cite{kung2008dihydroxylphenyl, murray2010fragment, stebbins1997crystal, janin2010atpase, li2019heat, yoshimura2021thermodynamic}.

Additionally, we found that some scaffolds of molecules with measurable binding affinities shown in Figure~\ref{fig:exp}\textbf{B} have been validated by previous studies. For instance, compounds 2, 7, 10, and 11 were inhibitors containing the resorcinol-like scaffold discovered in natural products. Compounds 5 and 9 shared a similar backbone belonging to the purine scaffold, while compound 8 was similar to an aminotriazine compound previously reported in a fragment-based screening~\cite{li2019heat, huth2007discovery}. Notably, we found that compounds 1, 3, 4, 6, and 12 have the potential to serve as promising novel candidates for further hit-to-lead optimization. To the best of our knowledge, there is no known precedent in the literature for these compounds, making them particularly valuable for future lead optimization. Quantitative analysis further showed that the 20 compounds identified in the virtual screening have low similarity compared to the HSP90 ligands in CASF-2016, with a Tanimoto similarity of only 0.270 for the ECFP4 fingerprints of the most similar molecule pair (Figure~S4, Supporting Information). In summary, our wet-lab validation results demonstrated that Dynaformer can deliver effective hit compounds with favorable binding affinities and novel scaffolds, thus facilitating the drug discovery process.

\section{Discussion}\label{sec3}

Protein-ligand binding affinity prediction is a fundamental problem in early drug discovery, as it can greatly help reduce the immense costs associated with wet-lab experiments in lead compound design. In this study, we first curated a comprehensive MD trajectory dataset consisting of 3,218 complexes containing proteins and bound small molecules. Each trajectory, with conformation snapshots from the thermodynamic ensemble, approximates the conformation space distribution and represents the ligand binding free energy state. We then developed Dynaformer, a graph transformer model that learns the underlying physico-chemical patterns of ligand binding from this MD trajectory dataset. Through extensive testing and ablation studies, we showed that MD trajectories contain rich features related to enthalpy and entropy and thus provide more information than static crystal structures. Our results demonstrated that Dynaformer significantly outperformed baseline approaches on the CASF-2016 benchmark dataset, highlighting the importance of incorporating MD trajectory data to enhance the scoring and ranking power of binding affinity prediction. We further illustrated the efficacy of Dynaformer as a scoring function in a real-world hit discovery process against HSP90. By scoring the docked poses of a library containing 50,000 molecules, Dynaformer can effectively identify hit compounds with favorable binding affinities and novel scaffolds. Among the 20 experimentally tested compounds, 12 molecules exhibited measurable binding affinities, including two compounds with submicromolar $K_i$ values.

Our work underscores the potential of employing MD trajectories and deep learning models to enhance binding affinity prediction. In both a benchmark dataset and a real-world drug discovery scenario, we have illustrated the effectiveness of Dynaformer as a scoring function and its ability to discover novel hit compounds. We anticipate that prediction performance can be further improved by incorporating more high-quality data and carefully designed training tasks.

\section{Experimental Section}\label{sec4}

\subsection{Decomposition of the Free Energy Related to the Protein-Ligand Binding Affinity}

The free energy $\Delta G$ of protein-ligand binding, which determines the binding affinity, could be decomposed into multiple components. Each component could be represented as a function associated with atomic features from the MD trajectories, providing a solid theoretical basis for learning structure-to-affinity relationships using deep learning models. Generally, $\Delta G$ could be expressed as the sum of two components, the gas-phase binding free energy and the solvation free energy, that is,
\begin{equation}\label{eq:affinity}
\begin{split}
    -RT\textrm{ln}K_i &= \Delta G = \Delta G_{gas} + \Delta G_{solv}
\end{split}
\end{equation}
where $\Delta G_{gas}$ represents the interaction energy between the given pair of protein and ligand at their gas phase, and  $\Delta G_{solv}$ stands for the solvation free energy difference between the protein-ligand complex and the sum of stand-alone protein and ligand systems. $R$ and $T$ stand for the universal gas constant and temperature, respectively, and $K_i$ stands for the thermodynamic equilibrium constant of the binding process.

The energy term $\Delta G_{gas}$ is the binding free energy between protein and ligand in the gas phase, and it can be further split into two parts, i.e., the enthalpy term $\Delta H_{gas}$ and the entropy term $\Delta S_{gas}$. Following the calculation scheme of the interaction entropy method~\cite{duan2016interaction, sun2017interaction}, both $\Delta H_{gas}$ and $\Delta S_{gas}$ can also be expressed with the interaction energies $E^{int}$, that is, 
\begin{equation}\label{eq:ggas}
\begin{split}
    \Delta G_{gas} &= \Delta H_{gas} - T\Delta S_{gas}\\
    &= \langle E^{int} \rangle + RT\textrm{ln} \langle e^{ \beta \Delta E^{int}} \rangle 
\end{split}
\end{equation}
where the enthalpy $\Delta H_{gas} $ and entropy $-T\Delta S_{gas} $ terms correspond to $ \langle E_{pl}^{bind} \rangle$ and $ RT\textrm{ln} \langle e^{ \beta \Delta E^{int}} \rangle$, respectively, $\beta$ stands for a constant $ 1/RT$, $\langle E^{int} \rangle$ stands for the trajectory-averaged interaction energy of protein and ligand contributed by different interaction types, such as hydrogen and hydrophobic interactions, Van der Waals interactions, $\pi$-$\pi$ stackings, and other long-range interactions. More specifically, $E^{int}$ could be expressed as $\sum\sum u(i, j)$, which was the sum of the interaction energy between each pair of atoms $i$ and $j$ from the protein and ligand, respectively. Here, $\Delta E^{int} = E^{int} - \langle E^{int} \rangle$ represents the fluctuation of the interaction energy around the trajectory-averaged value. 

The $\Delta G_{solv}$ term is the solvation free energy difference between the protein-ligand complex and the sum of stand-alone protein and ligand systems. It can be further split into two parts, i.e., the non-polar solvation free energy $\Delta G_{np}$ and the polar solvation free energy $\Delta G_{pol}$, that is,
\begin{equation}\label{eq:gsolv}
\begin{split}
    \Delta G_{solv}&= \Delta G_{np} + \Delta G_{pol}, \\
    &= \langle \sum \left( \gamma \mbox{SASA}(i) + \delta \right) \rangle + \langle \sum q_i \Phi(i) \rangle
\end{split}
\end{equation}
where the non-polar solvation free energy $\Delta G_{np}$ term, can be obtained using an empirical solvent-accessible surface area (SASA) formula parameterized by $\gamma$ and $\delta$, and $\Delta \mbox{SASA}$ stands for the change of solvent accessible surface area (SASA) upon binding~\cite{huang2018fast}. The polar solvation free energy $\Delta G_{pol}$ term was controlled by the electrostatic potential $\Phi(i)$, which could be obtained through solving the Poisson-Boltzmann (PB) equation or the Generalized Born (GB) equation~\cite{massova2000combined}. The $\Delta G_{pol}$ can then be calculated by summing over the electrostatic potential $\Phi(i)$ at the position of each atom $i$, weighted by the partial charge $q_i$ of atom $i$. 

In summary, each of the above components of the binding free energy change could be expressed as a function of the features of the atoms or atom pairs involved in the ligand binding. Therefore, they could be naturally represented and approximated through training the Dynaformer on the large-scale MD trajectory data. Since each term was actually averaged over the MD trajectory, the results could be more robust compared to those obtained from a single static crystal structure.

\subsection{Preparation of the MD Trajectory Dataset}\label{subsec2}

The protein-ligand complexes from the PDBBind dataset~\cite{wang2004pdbbind} were used as the initial conformations for the molecular dynamics simulations. PDBBind contained a comprehensive collection of high-quality crystal complex structures of proteins and ligands curated from the Protein Data Bank (PDB)~\cite{berman2000protein}, together with the corresponding experimentally measured binding affinity data. More specifically, the PDBBind dataset consisted of three parts, i.e., general, refined, and core sets. The general set included 21,382 complexes, while the refined set was a subset of the general set containing 4,852 structures, selected by a set of rules based on the quality of crystal structures and binding data~\cite{liu2017forging}. The core set, which was separately organized as CASF (Comparative Assessment of Scoring Functions), was a stand-alone test set consisting of 285 high-quality complexes covering 57 classes of targets for evaluating different computational models. The general and refined sets were updated annually, while the CASF set was not. Here, the model and all baseline models were evaluated on the latest version of CASF, i.e., CASF-2016.

Subsequently, the web-based CHARMM-GUI was used to clean up the data and prepare their corresponding MD simulation input files~\cite{jo2008charmm,lee2016charmm}. The MD simulations were performed based on the CHARMM36 force field~\cite{lee2016charmm} and using the NAMD engine, which was a parallel molecular dynamics program designed for high-performance simulation of large biomolecular systems~\cite{phillips2020scalable}. For each MD simulation, the complex was solvated in a truncated periodic TIP3P water box, and the minimum distance from the surfaces of the box to the complex atoms was set to 10 \r{A}. Counterions were added to neutralize systems, and the initial configuration was decided using a short Monte Carlo simulation. The ligand topology and parameter file were generated using the ParamChem service~\cite{ghosh2011molecular}. The missing residue was modeled using GalaxyFill~\cite{coutsias2004kinematic}. The simulation temperature was maintained at 303.15 K. For a stable protein-ligand complex, several nanoseconds were usually sufficient for free energy calculation, as in MM/PB(GB)SA~\cite{genheden2010obtain}. However, severe structural stability issues, such as ligand drifting and protein unfolding, become more obvious if the MD simulation duration was longer. Therefore, to tradeoff simulation time and computational complexity, a simulation time of 10 ns was finally chosen. A 0.5 ns NVT (constant volume and temperature) was set for the equilibrating stage before a 10 ns NPT (constant pressure and temperature) simulation was performed. 

In practice, any complex associated with the following conditions, which led to failed MD simulations was concluded: 1) covalent ligands that had a covalent bond with the protein; 2) complicated or multiple ligands that led to failed force field parametrization; 3) the protein structure contained too many missing residues and failed to repair; and 4) specific types of proteins, such as membrane proteins, which were not suitable for simulation in a water box. After removing the above cases, 3,218 protein-ligand complexes were selected from the PDBBind refined set. For each complex, a 10 ns simulation was performed, using the previously mentioned protocol, and from each simulation, 100 snapshots were sampled to form a trajectory.

\subsection{Data Preprocessing for Dynaformer}\label{subsec3}

In our study, we converted the structure of a protein-ligand complex, obtained from either an MD trajectory or a crystal structure, into a graph representation, which was subsequently fed into Dynaformer as input data. More specifically, we denote the ligand, the binding pocket, and their interatomic relationships as a graph, i.e., $\mathcal{G} = (\mathcal{V}, \mathcal{E})$, as depicted in Figure~\ref{fig:intro}\textbf{B}. Here, the node set $\mathcal{V} = \left\{ \mathbf{v}_0, \mathbf{v}_1, \mathbf{v}_2, \dots, \mathbf{v}_n \right\}$ represents the set of input atom feature vectors, where $n$ stands for the number of atoms and $\mathbf{v}_0$ refers to a virtual node employed as a global representation of the graph $\mathcal{G}$. The graph $\mathcal{G}$ is comprised of atom features derived from the ligand and the binding pocket. The binding pocket here is characterized by atoms that belong to the protein and are situated within a specific distance ($d_p$) from any atom of the ligand. The edge set $\mathcal{E} = \left\{\mathbf{e}_{i,j}~\vert~i,j \in \mathcal{V}\right\}$ stands for the set of covalent or non-covalent bond feature vectors, with $\mathbf{e}_{i,j}$ indicating the bond feature vector between atoms $i$ and $j$. Here, covalent interactions occur between atoms possessing chemical bonds, whereas non-covalent interactions transpire between any two atoms within a distance threshold ($d_s$).

We adopt the same features as used in the Open Graph Benchmark (OGB)~\cite{hu2020open}, and the comprehensive definitions of node and edge features can be found in Supplementary Table~S1. We extract the snapshots from MD trajectories using the MDAnalysis library~\cite{michaud2011mdanalysis}, and feature extraction is performed with the RDKit library~\cite{landrum2013rdkit}. In our experiments, we set $d_p = d_s = 5$\r{A}.

\subsection{The Model Architecture of Dynaformer}\label{subsec4}

Dynaformer was a transformer-based model that learned protein-ligand binding features from molecular dynamics data. Building on Graphormer~\cite{ying2021transformers}, a roto-translation invariant encoding scheme was proposed to encode the atomic features. At a high level, the Dynaformer predicted the binding affinity given the collective interaction patterns of a ligand at the active binding site. 

First, how Dynaformer encodes the distance and angle features between atoms was highlighted. As previously stated, binding free energy depends on the atomic features and interactions of atom pairs in the complex. To represent these scalar values for distances and angles, a Gaussian basis function (GBF) was utilized. The GBF was advantageous in modeling non-linear relationships between atom pairs, such as Van der Waals and electrostatic potentials~\cite{cisneros2012application}. Moreover, the GBF was a smooth function that could be seamlessly integrated into neural networks. The specific calculations for structural encodings are performed using the subsequent formulas:
\begin{equation}\label{eq:gbf}
\begin{aligned}
    \mbox{GBF}(x_k) &= \exp\left[-\frac{(x_k-\mu_k)^2}{2\sigma_k^2}\right], k=1,\dots,K\\
    d(i, j) &= \mbox{GBF}\left(\left[ \mathbf{v}_i \vert r_{ij} \vert \mathbf{v}_j \right]\mathbf{W}_d\right)\\
    a(i, j) &= \mbox{GBF}\left(\left[ \mathbf{v}_i \vert \sum_k\angle{ijk} \vert \mathbf{v}_j \right]\mathbf{W}_a\right)
\end{aligned}
\end{equation}
where $\mu_k$ and $\sigma_k$ stand for the learnable parameters, $K$ stands for the number of encoding heads, $x_k$ stands for a distance or angle value for encoding, $\mathbf{v}_i$ stands for the atomic features of atom $i$, $~\vert$ stands for the concatenation operation between vectors, $r_{ij}$ represents the Euclidean distance scalar in \r{A}s between atoms $i$ and $j$, $\angle{ijk}$ indicates the angle scalar in degrees between bonds formed by atoms $i$, $j$, and $k$, and $\mathbf{W}_d$, $\mathbf{W}_a$ correspond to the weight matrices for distance and angle features, respectively.

Then, the Dynaformer architecture was briefly introduced and emphasize the self-attention module. Dynaformer had two main components in each encoder layer: a multi-head self-attention module (MHA) and a feed-forward network (FFN). The encodings of structural features as described in Figure~\ref{fig:intro}\textbf{C} are fused into an MHA module, which then updates the feature representations of each atom and their contributions to the final binding affinity. In each head of MHA, the attention is calculated as follows:
\begin{gather}
    \mbox{Attention}(\mathbf{Q}, \mathbf{K}, \mathbf{V}) = \mbox{softmax}\left( \mathbf{A} \right)\cdot \mathbf{V}, \label{eq:single-attn} \\
    A_{ij} =\frac{\left(\mathbf{h}_i\mathbf{W_Q}\right)\left(\mathbf{h}_j \mathbf{W_K}\right)^T}{\sqrt{d_K}} + d(i,j) + a(i,j) + \frac{1}{P}\sum_{p=1}^{P} \mathbf{e}_p \mathbf{w}_p^T, \label{eq:dyna-attn}
\end{gather}
where $\mathbf{Q} = \mathbf{H}\mathbf{W_Q}, \mathbf{K}=\mathbf{HW_K}, \mathbf{V}=\mathbf{HW_V}$ stand for the query, key, and value matrices in the transformer model, $\mathbf{H} = \{\mathbf{h}_i\}$ stands for hidden representations of nodes, matrix $\mathbf{A}$ stands for the self-attention matrix, $A_{ij}$ is the unnormalized attention score between atoms $i$ and $j$, and $d_K$ stands for the dimensionality of the query, key, and value, which serves as a normalization term. In addition, the term $\frac{1}{P}\sum_{p=1}^{P} \mathbf{e}_p \mathbf{w}_p^T$ encodes the interatomic interactions on the shortest path between atoms $i$ and $j$, where $\mathbf{w}_p$ denotes the weight that projects edge feature vector $\mathbf{e}_p$ to a scalar and $P$ stands for the maximum length of the shortest path to keep the calculation feasible. This term is an extension of the adjacency matrix, which encodes the topological relationships between any pair of atoms.

In this MHA module, the feature representations of nodes were updated through the attention matrix $\mathbf{A}$, which was a weighted summation of all node representations. In every layer of Dynaformer, the atomic representations gradually exchange information and aggregate into the representation of the virtual node. After the last encoder layer, the feature representation of the virtual node was fused with the three pre-calculated fingerprints from ECIF~\cite{sanchez2021extended}, RF-score~\cite{ballester2010machine}, and GB-score~\cite{rayka_firouzi_2022} for predicting the binding affinities. Such a scheme allowed the Dynaformer to learn the effective structural patterns contributing to the final binding affinities. A detailed explanation of the model architecture is stated in the Supporting Information. In summary, the Dynaformer model offered a novel approach to predict protein-ligand binding affinities by utilizing a roto-translation invariant encoding scheme for atomic and structural features.

\subsection{Training and Hyperparameter Settings}\label{sec_data}

The training of Dynaformer consisted of two stages, i.e., pretraining and finetuning. The model configuration and the pretraining hyperparameters are summarized in Table~S2 (Supporting Information). The binding affinities of MD trajectories were labeled according to their PDB IDs, which means that all snapshots of one trajectory share the same label. The mean squared error (MSE) loss function was used to pretrain the model. Different from the vanilla transformer~\cite{vaswani2017}, four layers with 512 hidden dimensions were chosen through grid searching. An adversarial graph data augmentation method, named FLAG~\cite{kong2020flag}, was also used to mitigate the potential overfitting problem in the pretraining stage. For the finetuning stage, Dynaformer was finetuned on the PDBBind general set. In this stage, the mean absolute error (MAE) was used as a loss function as it resulted in larger loss values. In addition, the peak learning rate was adjusted to $10^{-5}$ and the batch size was set to 16. At both stages, the datasets were split into training and validation subsets with a ratio of 9:1. The validation set was used to guide the training or finetuning process. The training process was taken in about 24 h with eight NVIDIA V100-32GB GPUs.

\subsection{Hit Discovery Details Against HSP90}
The HSP90 protein structure and small molecules from a compound library were prepared for docking. The protein structure and the corresponding pocket position were determined from a reference structure (PDB ID: 2xdl) from the RCSB PDB. PyMOL~\cite{PyMOL} (version 2.5.0, open-source) was used to remove the solvent molecules, ions, and the original ligand. The average coordinates of the original ligand, $x=65.4, y=35.0, z=25.5$, were utilized as the center of the docking box. Hydrogens and charges were added to the protein and converted the conformation into the PDBQT format using ADFRsuite~\cite{ravindranath2015autodockfr} (version 1.0, build 5). For small molecules, the ETKDG algorithm from the RDKit library~\cite{landrum2013rdkit} (version 2022.3.3) was used to generate the initial conformations of molecules from the Chembridge DIVERSet-EXP library. Up to ten conformers were generated for each molecule, with a minimum heavy atom $\mbox{RMSD} = 0.5$\r{A} between conformers. Here, redocking could help stride over the energy barrier and thus reach the stable docking pose. All initial conformers were processed using the Meeko Python library (https://github.com/forlilab/Meeko), including salt removal, hydrogen addition, and partial charge assignment. Finally, all initial conformers were converted to the PDBQT format before docking. Then, Autodock Vina~\cite{eberhardt2021autodock} (version 1.2.3) was utilized for docking the protein and small molecule conformers. The docking box size was chosen to be 25\r{A}, and the exhaustiveness of the global search was set to 32. To better examine the capability of Dynaformer, the most basic molecular docking setup was used, in which flexible docking was not taken into account. Autodock Vina was run in parallel on 3200 CPUs, and all docking jobs were finished within 20 h. The best docking pose and the corresponding docking energy of each molecule were kept. The pipeline of receptor preparation, ligand preparation, and docking is shown in Figure~S1 (Supporting Information).

Subsequently, the docked structures undergo binding affinity prediction using Dynaformer within one hour. The top 10\% of molecules from both Dynaformer predictions and Autodock Vina docking energy were combined together. Following the literature~\cite{lyu2019ultra, bender2021practical}, a filtering and visual inspection procedure was conducted. In particular, a filtering process according to Veber's rule and Lipinski's rule of five~\cite{lipinski1997experimental, veber2002molecular} was used. More specifically, molecules were removed that have more than five hydrogen bond donors, more than ten hydrogen bond acceptors, a calculated octanol-water partition coefficient (Clog P) exceeding five, more than ten rotatable bonds, or a polar surface area larger than 140 $\mbox{\r{A}}^2$. Additionally, molecules with high intrinsic energy, such as those containing 3-, 4-, 7-, or more than 12-membered rings, were excluded. Then, a visual inspection was performed based on the HSP90 ATP binding pocket. Following similar principles in previous structure-activity analysis, promising molecules were selected based on whether the ligand possesses a hydrophobic head and a hydrophilic tail~\cite{stebbins1997crystal, huth2007discovery, kung2008dihydroxylphenyl, murray2010fragment, janin2010atpase,yoshimura2021thermodynamic}. Up to 20 molecules were selected for the subsequent wet lab validation experiments (Figure~S2, Supporting Information).

Finally, the binding affinities of the above 20 molecules were experimentally validated. Surface plasmon resonance (SPR) measurements were performed by the Biacore S200 instrument (GE Healthcare Life Sciences). HSP90 (9-236) was cross-linked on the surface of the Series S Sensor Chip CM5 (GE Healthcare) at pH 4.0 by an amine coupling kit with 1-ethyl-(3-dimethylaminopropyl)-carbodiimide hydrochloride (EDC) and NHS. Experiments were performed in a running buffer (1 $\times$ PBS with 5\% DMSO). Each compound was serially diluted by a factor of two in the running buffer. The compound solution flowed through the surface of the chip at a flow rate of 30 \textmu{}L/min. Sensorgrams were analyzed and generated by Biacore Evaluation Software. The inhibition constants $K_i$ of 8 out of 20 molecules could not be determined, probably due to poor aqueous solubility or non-specific binding.

\subsection{Binding Affinity Evaluation Metrics}



\noindent\textbf{Pearson $\bm{r}$.} The Pearson correlation coefficient, also known as the Pearson $\bm{r}$, measures the linear relationship between two sequences of data points: $\mathbf{x}=\{x_i\}$ and $\mathbf{y}=\{y_i\}$. More specifically, it can be defined as:
\begin{equation}
    \bm{r}(\mathbf{x}, \mathbf{y}) = \frac{\sum_{i=1}^{N}(x_i - \overline{x})(y_i - \overline{y})}{\sqrt{\sum_{i=1}^{N}(x_i - \overline{x})^2\sum_{i=1}^{N}(y_i - \overline{y})^2}}
\end{equation}
where $N$ stands for the number of samples, and $\overline{x}, \overline{y}$ stand for the mean values of $\mathbf{x}=\{x_i\}$ and $\mathbf{y}=\{y_i\}$, respectively. Pearson $\bm{r}$ varies between $[-1, 1]$, $0$ indicates no correlation, and $-1$ or $1$ indicate the exact negative or positive linear relationship. Note that Pearson $\bm{r}$ is invariant to the scale or location of the two variables, which means that it did not suggest the error between prediction and groundtruth.

\noindent\textbf{RMSE.} The root mean squared error (RMSE) was used to determine the average error of binding affinity prediction, as a complementary evaluation towards Pearson $\bm{r}$. It is defined as:

\begin{equation}
    \mbox{RMSE}(\mathbf{x}, \mathbf{y}) = \sqrt{\frac{1}{N}\sum_{i=1}^N (x_i - y_i)^2}
\end{equation}
where $N$ is the number of samples, $\mathbf{x}=\{x_i\}$ and $\mathbf{y}=\{y_i\}$ are the predictions and the groundtruth binding affinity values.




\noindent\textbf{Spearman $\bm{\rho}$.} The Spearman correlation coefficient, denoted as Spearman $\bm{\rho}$ is a nonparametric measure of the ranking order between two sequences of data points. In other words, Spearman $\bm{\rho}$ has a similar form as Pearson $\bm{r}$ but calculates the ranking values $\mathbf{rx}$ and $\mathbf{ry}$ to assess the monotonic relationships between variables. More specifically, it can be defined as:
\begin{equation}
    \bm{\rho}(\mathbf{x}, \mathbf{y}) = \frac{\sum_{i=1}^{N}(rx_i - \overline{rx})(ry_i - \overline{ry})}{\sqrt{\sum_{i=1}^{N}(rx_i - \overline{rx})^2\sum_{i=1}^{N}(ry_i - \overline{ry})^2}}
\end{equation}
where $N$ stands for the number of samples, $\mathbf{rx}=\{rx_i\}$ and $\mathbf{ry}=\{ry_i\}$ stand for the ranking orders of complexes between the predicted and experimental binding affinities, respectively, and $\overline{rx}, \overline{ry}$ stand for the mean rank orders, respectively. Spearman $\bm{\rho}$ is invariant to the specific values, but it is sensitive to the order of variables.

\noindent\textbf{Kendall $\bm{\tau}$.} The Kendall correlation coefficient, denoted as Kendall $\bm{\tau}$, is a measure of the ordinal association between two sequences of ranking data points. More specifically, it can be defined as:
\begin{equation}
    \bm{\tau}(\mathbf{x}, \mathbf{y}) = \frac{N_{concord}-N_{discord}}{\sqrt{(N_{concord}+N_{discord}+\mathcal{T})(N_{concord}+N_{discord}+\mathcal{U})}}
\end{equation}
where $N_{concord}$ and $N_{discord}$ stand for the numbers of concordant and discordant pairs, respectively. Here, given two pairs of prediction and groundtruth $\{(x_1, y_1), (x_2, y_2)\}$, if $\mbox{sign}(x_2-x_1) = \mbox{sign}(y_2-y_1)$, this pair is called a concordant pair. In contrast, if $\mbox{sign}(x_2-x_1) = -\mbox{sign}(y_2-y_1)$, this pair is called a discordant pair. A pair is tied if $x_1=x_2$ or $y_1=y_2$. $\mathcal{T}$ and $\mathcal{U}$ are the numbers of tied pairs in $\mathbf{x}$ and $\mathbf{y}$, respectively. Kendall $\bm{\tau}$ is invariant to the specific values, but it is sensitive to the order of variables.

\noindent\textbf{PI.} The predictive index (PI) is an evaluation metric mainly proposed for virtual screening~\cite{pearlman2001free}. In particular, PI measures how well a scoring function can correctly rank different ligands for a specific target by taking the differences in binding affinities into account. More specifically, it can be defined as:
\begin{equation}
    \mbox{PI}(\mathbf{x}, \mathbf{y}) = \frac{\sum_{j>i}^N\sum_{i=1}^N\omega_{ij}S_{ij}}{\sum_{j>i}^N\sum_{i=1}^N\omega_{ij}}
\end{equation}
where $\omega_{ij}$ stands for the absolute difference of binding affinities between ligands $i$ and $j$, i.e., $\omega_{ij} = \vert\mbox{p}K_i^{(i)} - \mbox{p}K_i^{(j)}\vert = \vert y_i - y_j \vert$, $S_{ij} \in \{1, 0\}$ indicates if the ranking orders of binding affinities xi and xj are consistent between prediction and groundtruth.

\subsection{Molecular Structure Analysis Metrics}
\textbf{RMSD.} The root mean squared deviation (RMSD) measures the average deviation of a matching set of atoms between two structures. It is expressed as:

\begin{equation}
    \mbox{RMSD} = \sqrt{ \frac{1}{N} \sum^N_{i=1} \delta^2_i}
\end{equation}
where $N$ stands for the number of samples and $\delta_i$ stands for the Euclidean distance between the atom $i$ and the reference position, i.e., usually the initial position.

\noindent\textbf{RMSF.} The root mean squared fluctuation (RMSF) was the fluctuation around an average, per atom or residue, over a sequence of structures, such as a trajectory. More specifically, it can be expressed as:
\begin{equation}
    \mbox{RMSF} = \sqrt{\langle(x_i - \overline{x_i})^2\rangle}
\end{equation}
where $x_i$ stands for the coordinates of the particle $i$, and $\overline{x_i}$ stands for the trajectory-averaged position of $i$. 

The RMSD quantified how much a structure diverges from a reference over time, while the RMSF could measure the mobility of parts of the system. Thus, parts of the structure with high RMSF values diverge from the average, indicating high mobility.


\subsection{Statistical Analysis}
The binding affinity values $K_i$ were preprocessed into $\mbox{p}K_i = -\log_{10}K_i$. The performance of binding affinity prediction was evaluated using multiple metrics: Pearson correlation coefficient, root-mean-square error (RMSE), standard deviation (SD), Spearman correlation coefficient, Kendall correlation coefficient, and predictive index (PI). Statistical significance was assessed using appropriate multiple comparison tests. P-values were adjusted for multiple hypothesis testing as specified in the text. Confidence intervals (CIs) were calculated to estimate the precision of the obtained results. Data were presented as means and standard deviations. All statistical analyzes were performed using Python 3.10 with NumPy, SciPy, MDAnalysis, and Seaborn packages. Protein-ligand complex structures were visualized using PyMol 2.4 (open-source version). Small molecule structures were visualized with ChemSketch freeware.

\section{Data Availability}
The data that support the findings of this study are openly available in OneDrive at at \url{https://1drv.ms/f/s!Ah9r82oejjV8piQHq_qAieio_86z?e=B1E53d}. The repository contains MD trajectories, scripts for MD simulation and preprocessed data for Dynaformer.

\section{Code Availability}

The Dynoformer is implemented in Python. The source code, trained models as well as evaluation results are available at \url{https://github.com/Minys233/Dynaformer}.

\section*{Supplementary Information}\label{sec_suppl}
The supplementary material is available in the attachment.

\section*{Acknowledgments}

This work was supported in part by the National Natural Science Foundation of China (T2125007 to JZ, 32270640 to DZ), the National Key Research and Development Program of China (2021YFF1201300 to JZ), the New Cornerstone Science Foundation through the XPLORR PRIZE (JZ), the Turing AI Institute of Nanjing, the Research Center for Industries of the Future (RCIF) at Westlake University (JZ) and the Westlake Education Foundation (JZ), and "Pioneer" and "Leading Goose" R\&D Program of Zhejiang (2024SSYS0036). The authors thank Dr. Hao Zhang and Dr. Tristan Bereau for the insightful discussions.

\section*{Author Contributions}

Y. M., Y. W., and P. W. are contributed equally to this work as co-first authors. Y.W., D.Z., and J.Z conceived the idea; J.Z. and J.W. planned the dry lab and wet lab experiments; Y.W., P.W., and Y.M. performed MD simulations and quality assessment; Y.M., Y.W., P.W., J.Z., S.Z., and J.W. designed the Dynaformer model; Y.M., Y.S., and Y.H.W trained and tested the Dynaformer model; Y.M. conducted virtual screening; J.Z., D.Z., and X.W. directed SPR assays; Y.M., Y.W., J.Z., D.Z., and J.W. analyzed the data; Y.M., P.W., Y.W., H.L., and X.W. wrote the original drafts, and all authors commented on the manuscript.

\section*{Conflict of Interest}

The authors declare no conflict of interest.


\bibliographystyle{sn-standardnature} 
\bibliography{sn-article}


\newpage
\section*{Figure legends}

\newpage

\clearpage

\newpage

\newpage

\end{document}